\documentclass[prl,aps,twocolumn,tightenlines,showpacs,superscriptaddress]{revtex4}  
\usepackage{graphics}

\def\TMG{\tau\rightarrow\mu\gamma}
\def\TT{\tau\tau}
\def\GT{\tau}
\def\GG{\gamma}
\def\GM{\mu}
\def\MG{\mu\gamma}

\def\BB{{B\overline{B}}}

\def\Minv{M_{inv}}
\def\DelE{\Delta E}

\begin{document}

\title{\large
       An Upper Bound on the Decay $\TMG$ from Belle}

\affiliation{Budker Institute of Nuclear Physics, Novosibirsk}
\affiliation{Chiba University, Chiba}
\affiliation{University of Cincinnati, Cincinnati, Ohio 45221}
\affiliation{Gyeongsang National University, Chinju}
\affiliation{University of Hawaii, Honolulu, Hawaii 96822}
\affiliation{High Energy Accelerator Research Organization (KEK), Tsukuba}
\affiliation{Hiroshima Institute of Technology, Hiroshima}
\affiliation{Institute of High Energy Physics, Chinese Academy of Sciences, Beijing}
\affiliation{Institute of High Energy Physics, Vienna}
\affiliation{Institute for Theoretical and Experimental Physics, Moscow}
\affiliation{J. Stefan Institute, Ljubljana}
\affiliation{Kanagawa University, Yokohama}
\affiliation{Korea University, Seoul}
\affiliation{Kyungpook National University, Taegu}
\affiliation{Institut de Physique des Hautes \'Energies, Universit\'e de Lausanne, Lausanne}
\affiliation{University of Ljubljana, Ljubljana}
\affiliation{University of Maribor, Maribor}
\affiliation{University of Melbourne, Victoria}
\affiliation{Nagoya University, Nagoya}
\affiliation{Nara Women's University, Nara}
\affiliation{National Kaohsiung Normal University, Kaohsiung}
\affiliation{National Lien-Ho Institute of Technology, Miao Li}
\affiliation{Department of Physics, National Taiwan University, Taipei}
\affiliation{H. Niewodniczanski Institute of Nuclear Physics, Krakow}
\affiliation{Nihon Dental College, Niigata}
\affiliation{Niigata University, Niigata}
\affiliation{Osaka City University, Osaka}
\affiliation{Osaka University, Osaka}
\affiliation{Panjab University, Chandigarh}
\affiliation{Peking University, Beijing}
\affiliation{Princeton University, Princeton, New Jersey 08545}
\affiliation{University of Science and Technology of China, Hefei}
\affiliation{Seoul National University, Seoul}
\affiliation{Sungkyunkwan University, Suwon}
\affiliation{University of Sydney, Sydney NSW}
\affiliation{Tata Institute of Fundamental Research, Bombay}
\affiliation{Toho University, Funabashi}
\affiliation{Tohoku Gakuin University, Tagajo}
\affiliation{Tohoku University, Sendai}
\affiliation{Department of Physics, University of Tokyo, Tokyo}
\affiliation{Tokyo Institute of Technology, Tokyo}
\affiliation{Tokyo Metropolitan University, Tokyo}
\affiliation{Tokyo University of Agriculture and Technology, Tokyo}
\affiliation{Toyama National College of Maritime Technology, Toyama}
\affiliation{University of Tsukuba, Tsukuba}
\affiliation{Utkal University, Bhubaneswer}
\affiliation{Virginia Polytechnic Institute and State University, Blacksburg, Virginia 24061}
\affiliation{Yokkaichi University, Yokkaichi}
\affiliation{Yonsei University, Seoul}
  \author{K.~Abe}\affiliation{High Energy Accelerator Research Organization (KEK), Tsukuba} 
  \author{K.~Abe}\affiliation{Tohoku Gakuin University, Tagajo} 
  \author{T.~Abe}\affiliation{High Energy Accelerator Research Organization (KEK), Tsukuba} 
  \author{I.~Adachi}\affiliation{High Energy Accelerator Research Organization (KEK), Tsukuba} 
  \author{Byoung~Sup~Ahn}\affiliation{Korea University, Seoul} 
  \author{H.~Aihara}\affiliation{Department of Physics, University of Tokyo, Tokyo} 
  \author{K.~Akai}\affiliation{High Energy Accelerator Research Organization (KEK), Tsukuba} 
  \author{M.~Akatsu}\affiliation{Nagoya University, Nagoya} 
  \author{M.~Akemoto}\affiliation{High Energy Accelerator Research Organization (KEK), Tsukuba} 
  \author{Y.~Asano}\affiliation{University of Tsukuba, Tsukuba} 
% ...
\collaboration{The Belle Collaboration}

\begin{abstract}
We have performed a search for the lepton-flavor-violating decay 
$\TMG$ using a data sample of 86.3~fb$^{-1}$ accumulated by 
the Belle detector at KEK. 
No evidence for a signal is seen, and 
we set an upper limit for the branching fraction of
$\mathcal{B}(\TMG)<3.1\times 10^{-7}$ at the 90\% confidence level.  
\end{abstract}


\maketitle

The decay $\TMG$ violates lepton flavor conservation and
is forbidden within the Standard Model (SM).
However, some supersymmetric models, left-right symmetric 
models, and others~\cite{two} predict a
branching fraction in the range $10^{-7}$ to $10^{-9}$, which
is accessible at an $e^+e^-$ B-factory.
It is notable that $\mathcal{B}(\tau\rightarrow \mu\gamma)$ could be
enhanced by a factor of $10^5$--$10^6$ relative to 
$\mathcal{B}(\mu\rightarrow e\gamma)$, because relevant 
kinematical factors depend on powers of $m_{\tau}/m_{\mu}$. 
The decay $\TMG$ is thus a promising process in which to search for 
new physics. 

This decay has previously been searched for by MARK-II \cite{markII}, 
ARGUS \cite{argus}, DELPHI \cite{delphi}, 
CLEO \cite{CLEO1,CLEO}, and BaBar \cite{babar}. 
The most sensitive upper limit, reported by CLEO,
is $\mathcal{B}(\TMG) <$ $1.1\times 10^{-6}$ at 90\% C.L. 

We present here our study using 86.3 fb$^{-1}$ of 
data collected at the $\Upsilon(4S)$ resonance with the 
Belle detector at the KEKB asymmetric $e^+e^-$ collider \cite{kekb}.
A description of the detector can be found in Ref.~\cite{BelleDetector}.

We search for an event composed of exactly two oppositely-charged tracks 
and at least one photon candidate, 
which is consistent with a $\tau^+\tau^-$ event 
in which one $\tau$ decays to $\MG$ and the other $\tau$ decays to 
a non-muon charged particle, neutrino(s) and any number of $\GG$'s. 

The selection criteria are determined by examining 
Monte Carlo (MC) simulations for signal $\tau$-pair decay 
and background events (BG), such as generic $\tau$-pair decay 
($\tau^+\tau^-$), 
$q\bar{q}$ continuum, $\BB$, Bhabha, and 
$\mu^+\mu^-$ as well as two-photon processes.
The KORALB/TAUOLA \cite{tauola} and QQ \cite{qq} generators are used for 
event generation, and GEANT3 \cite{geant} is used to simulate the Belle 
detector.
The two-body decay $\TMG$ is initially assumed to have a uniform angular 
distribution in the $\GT$'s rest system; 
possible deviations from this are considered later.

The selection criteria are chosen to maximize the signal sensitivity.
Kinematic variables with a CM superscript are calculated 
in the center-of-mass frame; all other variables are calculated in the 
laboratory frame.
Each track is required to have momentum $p^{CM} < 4.5$~GeV/$c$ 
and momentum transverse to the $e^+$ beam
$p_t \geq 0.1$~GeV/$c$.
Both tracks are required to be within the acceptance
of a muon identification system (KLM): $-0.819 < \cos\theta < 0.906$,
where $\theta$ is the polar angle with respect to the $e^+$ beam direction.
Muon candidates are identified via a muon relative 
likelihood ${\cal{L}}_{\mu}$~\cite{muid}, which is based on the difference 
between the range calculated from the particle momentum and the range 
measured by the KLM. 
It also includes the $\chi^2$ formed from the KLM hit locations with 
respect to the extrapolated track. 
The charged track that forms
a $\TMG$ candidate is required to have $p > 1.0$~GeV/$c$ and
${\cal{L}}_{\mu}>0.95$.
The other track (on the ``tag-side'') is required to have
${\cal{L}}_{\mu} < 0.80$; i.e., we require that it not be a muon 
to reduce $e^+e^-\rightarrow\mu^+\mu^-\gamma$ background
($\gamma$ from initial-state radiation).
     The fraction of muons with ${\cal{L}}_{\mu} < 0.80$
     is denoted as $\eta$ and is measured to be (for a tag-side muon) 
     $(12 \pm 3)\%$. 

     Photon candidates, whose definition is given in Ref.~\cite{gamma},
are required to be within the acceptance of 
an electromagnetic calorimeter (ECL): $-0.866<\cos\theta_{\gamma}<0.906$. 
     The photon that forms a $\TMG$ candidate is required to have 
     an energy $E_{\GG}$ $>$ 0.5~GeV in order to avoid 
     a spurious combination of a low-energy $\GG$ with the muon. 

\begin{figure}
\centerline{\resizebox{4.0cm}{!}{\includegraphics{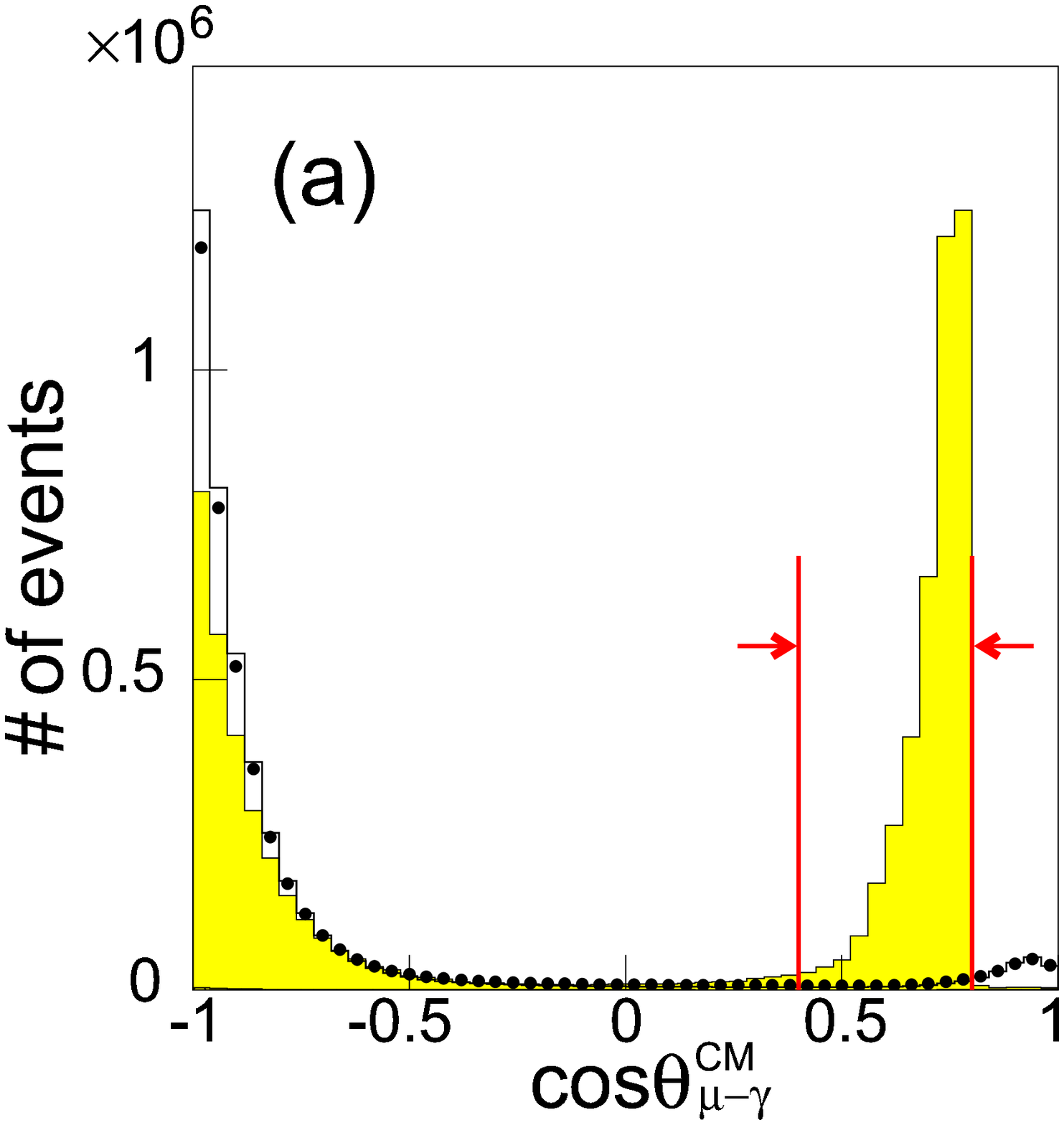}}
            \resizebox{4.0cm}{!}{\includegraphics{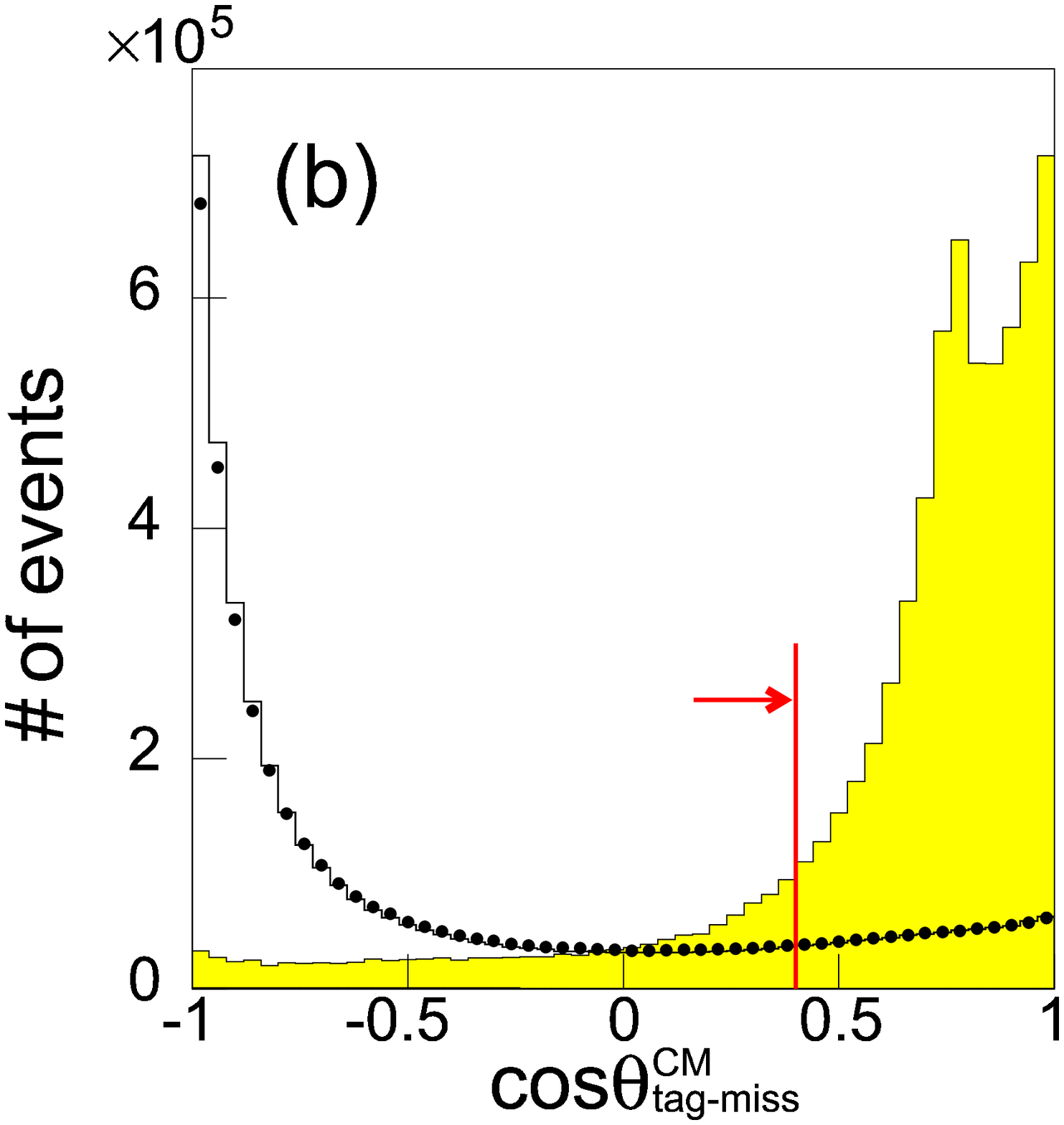}}}
\caption{Distributions of the opening angle 
	between (a) the $\mu$ and $\gamma$ 
        on the signal side, and (b) the tagging track and the missing momentum. 
	MC distributions for signal and $\tau^+\tau^-$ events
	are indicated by shaded and open histograms, respectively, 
	and data by closed circles. The arrows show the selected ranges.
}
\label{correl}
\end{figure}

A cut $E^{CM}_{sum} <$ 9.0~GeV is imposed to reject Bhabha 
	scattering and $\mu^+\mu^-$ production, where $E^{CM}_{sum}$ 
	is defined as the sum of the energies of the two charged tracks and 
	the photon composing the $\mu\gamma$.
	A restriction on the opening angle between the $\GM$ and $\GG$, 
	$0.4<\cos\theta_{\mu-\gamma}^{CM}<0.8$, is particularly powerful to
	reject background events arising mostly from 
	$e^+e^-\rightarrow\tau^+\tau^-$, $\tau \to \pi^0 X$.  
	This process forms the backward peak in
	the open histogram in Fig.~\ref{correl}(a). 
	The opening angle between the two tracks is required to be 
	greater than $90^\circ$. \par
We define $\vec{p}_{miss}$ as the residual momentum vector 
calculated by subtracting
the vector sum of all visible momenta (from tracks and photon candidates)
from the sum of the $e^+$ and $e^-$ beam momenta.
We include all photon candidates with energy greater than
0.1~GeV in the $p_{miss}$ calculation.
	Constraints on the momentum and polar angle of the missing particle(s), 
	$p_{miss} >$ 0.4~GeV/$c$ and $-0.866 < \cos\theta_{miss} < 0.956$, 
	respectively, 
	are imposed to increase the probability that the missing particle(s) 
	is an undetected neutrino(s) rather than $\GG$'s or charged 
	particles falling outside the acceptance of the detector.
To remove $\tau^+\tau^-$ events, we apply a requirement
to the opening angle between the tagging track and 
	the missing particle, $\cos\theta_{tag-miss}^{CM} >$ 0.4, 
as shown in Fig.~\ref{correl}(b). 

\begin{figure}
\centerline{\resizebox{4.3cm}{!}{\includegraphics{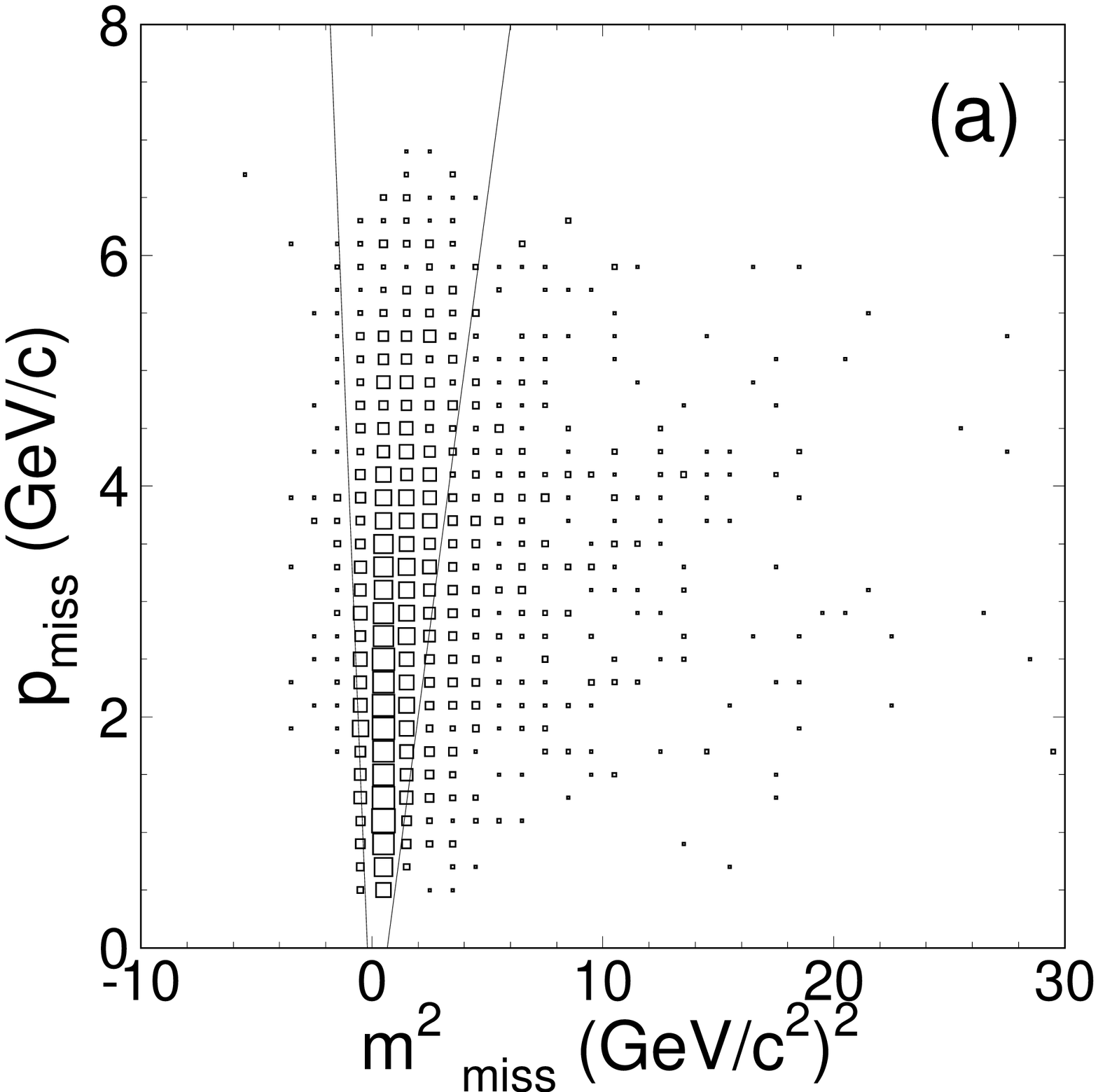}}
            \resizebox{4.3cm}{!}{\includegraphics{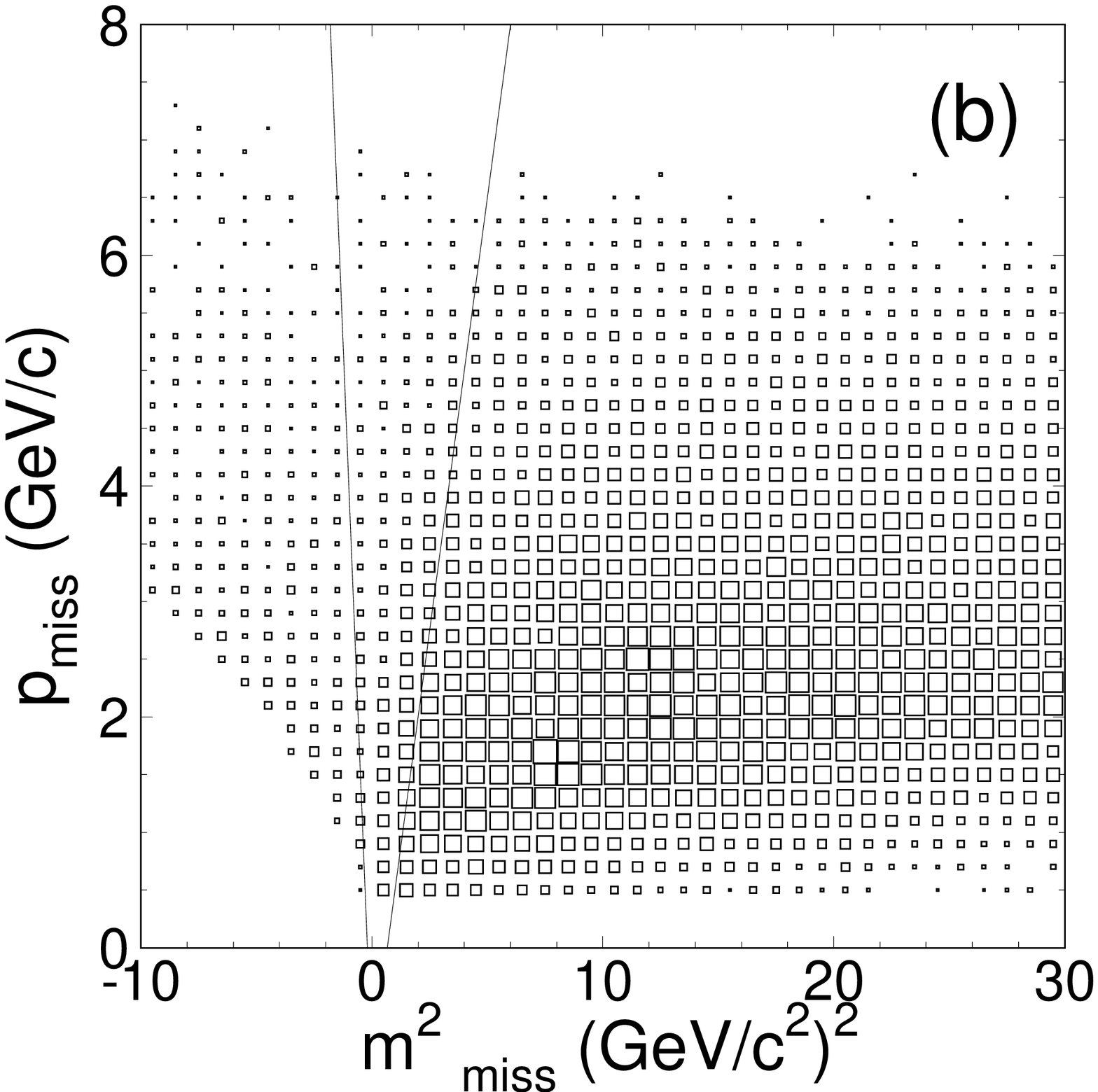}}}
\caption{Distribution of events in the $m_{miss}^2$-$p_{miss}$ plane.  
     The selection boundary is indicated by two lines 
     for (a) signal MC events and (b) $\tau^+\tau^-$ MC events.} 
\label{additi}
\end{figure}

Next, 
a condition is imposed on the relation between $p_{miss}$ and the mass-squared 
of a missing particle ($m_{miss}^2$).
		 The latter is defined as $E^2_{miss}-p^2_{miss}$, 
		where $E_{miss}$
                is 11.5~GeV (the sum of the beam energies)
		minus the sum of all visible energy
                and is calculated assuming
                the muon (pion) mass for the charged track on the
                signal (tag) side. We require
        $p_{miss} > -5 \times m_{miss}^2-1$ and
        $p_{miss} > 1.5 \times m_{miss}^2-1$,
	where $p_{miss}$ is in GeV/$c$ and $m_{miss}$ is in GeV/$c^2$.
	This cut loses 24\% of the signal but removes 98\% of 
	the remaining $\tau^+\tau^-$ background 
        (see Fig.~\ref{additi}) and 86\% of 
	the $\mu^+\mu^-$ background. 


After these selection requirements, 713 events remain in the data 
without any restriction on the mass and momentum of the $\mu\gamma$ 
system. 
The detection efficiency is
evaluated by MC to be $\epsilon=12.0\pm0.1$\%. 

\begin{figure}
\centerline{\resizebox{7.0cm}{!}{\includegraphics{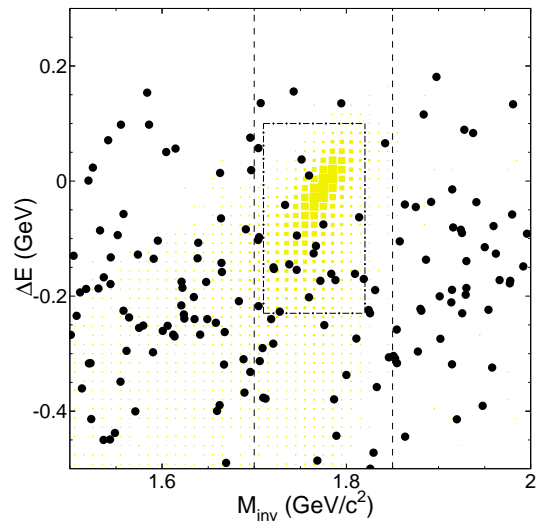}}}
\caption{
Remaining events in data (circles) and 
the expected density for signal MC (shaded)
in the $\DelE$-$\Minv$ plane. 
The region between
the dashed lines is kept excluded until the selection criteria are
finalized and the expected background estimated. The signal box
(defined in the text) is indicated by a dash-dotted box.
}
\label{Remain}
\end{figure}

The candidate $\mu\gamma$ system should have an invariant mass 
$(M_{inv})$ close to the $\tau$ lepton mass and 
an energy close to the beam energy in the CM frame; i.e., 
$\DelE=E^{CM}_{\mu\gamma}-E^{CM}_{beam}\simeq 0$. 
When deciding on our selection criteria, we excluded the signal region 
1.70~GeV/$c^2$ $< \Minv <$ 1.85~GeV/$c^2$ so as not to bias our choice 
of criteria (a ``blind'' analysis). 
Only after all cuts were finalized and the number of
background events estimated did we include 
this region and count the number of signal events. 
The resultant $\DelE$ vs. $\Minv$ plot is shown in Fig.~\ref{Remain}.
For comparison, the equivalent plot for MC $\TMG$  decays is also shown.
Because of the photon's energy leakage from the ECL detector 
and initial-state radiation, the MC simulation exhibits 
a long low-energy tail across the $\DelE$-$\Minv$ plane. 
The individual $\DelE$ and $\Minv$ distributions around the peak
are reproduced by asymmetric 
Gaussians with $\sigma_{\DelE}^{low/high}=(75.4\pm 0.7)/(33.7\pm 0.4)$ MeV
and $\sigma_{\Minv}^{low/high}=(23.1\pm 1.2)/(15.0\pm 0.6)$ MeV/$c^2$, 
where $\sigma^{low/high}$ means the standard deviation 
at the lower/higher side of the peak.  
The peak positions are $-1.1 \pm 0.5$ MeV and 1776.8$\pm$1.0 MeV/$c^2$ 
for $\DelE$ and $\Minv$, respectively.

We define a $\DelE$-$\Minv$ region to evaluate the number of signal candidates. 
The signal box is defined as the area within 
$\pm$3$\sigma$ for both $\DelE$ and $\Minv$: 
$-0.23$~GeV $< \DelE < 0.10$~GeV and 
$1.71$~GeV/$c^2$ $< \Minv < 1.82$~GeV/$c^2$.
The acceptance $(\Omega)$ for signal 
events passing all previous cuts is~87.3\%.

There are two dominant sources of background: 
$e^+e^- \rightarrow \mu^+\mu^- \gamma$ and 
$e^+e^- \rightarrow \tau^+\tau^- \gamma$, 
in which the photon is radiated from the initial state. 
In the former case the muon on the tag side is misidentified; 
in the latter case the muon on the signal side originates from 
$\tau\rightarrow\mu\nu\overline{\nu}$ decay. 
We hereafter denote the former background as $\mu\!\!\!/\mu \GG$ 
and the latter as $\TT\GG$. 
The $\TT\GG$ background is studied using a 150 fb$^{-1}$ sample of MC 
$\tau^+\tau^-$ events. 
The $\mu\!\!\!/\mu \GG$ background is studied using data by requiring 
that both tracks be muons
and applying the muon inefficiency ($\eta$)
to the tag-side track.
For this selection, the signal (tag)-side muon 
is required to have ${\cal{L}}_{\mu} > 0.95 (0.80)$.
Within the whole $\DelE$-$\Minv$ region shown in Fig.~\ref{Remain}, 
the $\TT\GG$ process yields 90.6$\pm$7.2 events and the $\mu\!\!\!/\mu \GG$
process 43.4$\pm$12.4 events. 
Among the other background processes mentioned above, only 
the continuum background yields a nonnegligible contribution
(12.7$\pm$5.7 events).
This background has a rather large uncertainty 
as it is evaluated using an MC sample corresponding to only 34~fb$^{-1}$.
The expected backgrounds amount to 143.7$\pm$15.5 events in total, where 
3.0 events in the $\mu\!\!\!/\mu \GG$ sample that 
are estimated to originate from 
$\TT\GG$ are subtracted to avoid double-counting. 
The number of data events in this region 
(now including the previously-excluded region) 
after all selection cuts is 160$\pm$13. 
This yield is consistent with the background estimate. 

The distributions of $\TT\GG$ and $\mu\!\!\!/\mu \GG$ background events, 
$N^{\TT\GG}(\Minv,\DelE)$ and $N^{\mu\!\!\!/\mu\GG}(\Minv,\DelE)$, 
exhibit different behavior in $\DelE$.
The former populates only the negative $\DelE$ region, while the latter 
is distributed mostly in the positive $\DelE$ region. 
Both types of background events, however, exhibit a similar correlation in
the $\Delta E$-$M_{inv}$ plane 
and are empirically reproduced by 
a combination of Landau and Gaussian functions 
in $\Delta E$, and a linear function in $M_{inv}$.
The background distribution can then be represented by the sum of two BG 
components as 
\begin{eqnarray}
N_{BG}(\Minv,\DelE)&=&
N^{\TT\GG}(\Minv,\DelE)\times (1+\Lambda) \nonumber\\
&+& N^{\mu\!\!\!/\mu \GG}(\Minv,\DelE)\times \kappa,  
\label{eq6}
\end{eqnarray}
where $\Lambda$ is the continuum background contribution.
This distribution is taken to be similar to that of $\TT\GG$,
as indicated by Monte Carlo simulation.
The factors $(1+\Lambda)$ and $\kappa\equiv\eta/(1-\eta)$ are
$1.14\pm 0.09$ and $0.14\pm 0.04$, respectively, 
where the former is obtained from MC and the latter is obtained 
by measuring the muon inefficiency $(\eta)$ for 
$e^+e^-\rightarrow\mu^+\mu^-$ events. 

\begin{figure}
\centerline{\resizebox{7.5cm}{!}{\includegraphics{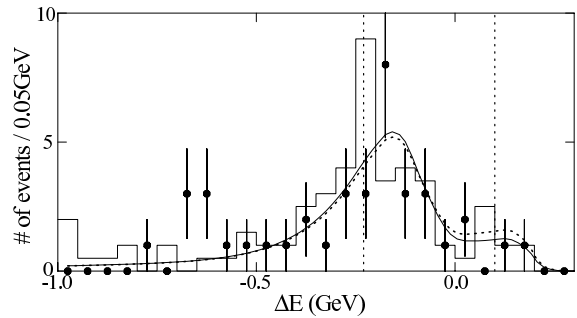}}}
\caption{$\DelE$ distributions in the signal region, 
         $1.71$~GeV/$c^2$ $<\Minv<1.82$~GeV/$c^2$. 
The expected background is indicated by 
	 the solid (dashed) curve for $1+\Lambda=1.22 (1.14)$ 
	 and $\kappa=0.11 (0.14)$.
	 The histogram is an average of both sideband spectra.
	 The data are the points with error bars.
	 The vertical dashed lines indicate the $\DelE$ range of the signal box.
}
\label{Compari}
\end{figure}

The above background parametrization is
compared with the data outside of the excluded region.
Fitting the background shape to the data gives 
$1+\Lambda=1.22\pm 0.13$ and $\kappa=0.11\pm 0.03$, which
agree well with the values given above. 
For both sets of $1+\Lambda$ and $\kappa$ values, 
Fig.~\ref{Compari} shows the expected background 
events $N_{BG}(\Minv,\DelE)$ as a function of $\DelE$ for 
$1.71$~GeV/$c^2$ $< \Minv <1.82$~GeV/$c^2$. 
No appreciable difference is seen between the two $N_{BG}(\Minv,\DelE)$
distributions, and 
we take the latter values for $1+\Lambda$ and $\kappa$. 
As the background has a linear (and in fact small) dependence on $\Minv$, 
the amount of background in the signal region can be estimated by 
averaging the number of events counted in the $\Minv$ sidebands 
$1.553 - 1.663$~GeV/$c^2$ and 
$1.850 - 1.960$~GeV/$c^2$. 
These sidebands begin a distance $2\sigma_{\Minv}$ from the edges 
of the signal region so that no potential signal biases this estimate. 
When the events in the signal box are included,
the data event distribution is found 
to match the expected $N_{BG}(\Minv,\DelE)$ well
(see Fig.~\ref{Compari}). 
Within the signal box, 19 events are found in the data while 
$20.2\pm2.1$ events 
are expected from eq.(\ref{eq6}) and $20.5\pm6.4$ events from the average
of $\Minv$ sidebands.

An upper limit is obtained using the frequentist method described in 
Ref.~\cite{CLEO, narsky}.
The likelihood function is defined as 
\begin{equation}
{\cal {L}}=\frac{e^{-(s+b)}} {N!} 
\prod_{i=1}^{N} (s S_i + b B_i) ,
\end{equation}
where $N$ is the number of observed events, and $s$ and $b$ are the number of 
signal events and background events, respectively. 
$S_i$ and $B_i$ are the signal and background probability density functions, 
where $i$ indicates the $i$-th event. 
$S_i$ is obtained by generating $100\times 10^6$ MC signal events, and 
$B_i$ is taken from eq.(\ref{eq6}).   
We apply this fit for $s$ and $b$ to a $\pm5\sigma$ region in $\DelE$ and $\Minv$,
which has an acceptance ($\Omega$) of 93.1\%. There are a total of
54 events in this region, and, when $s$ is constrained to be
$\geq$0, the fit finds $s=0.0$ and $b=54.0$.
The $\chi^2$ of the fit projection in $\DelE$ is 7.90 for 10 bins,
while in $\Minv$ it is 5.57 for 10 bins.
These values correspond to confidence levels (evaluated via toy MC) 
of 0.66 and 0.86, respectively.

To calculate the upper limit,
Monte Carlo samples are generated by fixing the 
expected number of background events (\~{b}) to the value $b=54$. 
For every assumed expected number of signal events (\~{s}), 10,000 samples are 
generated, for each of which the numbers of signal events and background 
events are 
determined by Poisson statistics with means of \~{s} and \~{b}, 
respectively. 
We then assign $\DelE$ and $\Minv$ values to these events 
according to their density distributions. 
An unbinned maximum likelihood fit is performed for 
every sample to extract the number of signal events $(s^{MC})$. 
The confidence level for an assumed \~{s} is defined as the fraction of 
the samples whose $s^{MC}$ exceeds $s$. 
This procedure is repeated until we find the value of \~{s} (\~{s}$_{90}$) 
that gives a 90\% chance of $s^{MC}$ being larger than $s$. 

The resulting upper limit at 90\% C.L. is \~{s}$_{90}=5.1$ events. 
An upper limit on the branching fraction 
is obtained via the formula: 
\begin{equation}
\mathcal{B}(\TMG) < \frac{\mbox{\~{s}}_{90}}{2(\epsilon\Omega) N_{\TT}}, \label{Beys0}
\end{equation}
where $N_{\TT}$ is the total number of $\tau$-pairs produced. 
Inserting the values \~{s}$_{90}=5.1$, $\epsilon=12.0\%$, $\Omega=93.1\%$
and $N_{\TT}=7.85\times 10^7$ gives
${\mathcal{B}}(\TMG) < 2.9 \times 10^{-7}$\footnotemark[2]. 

\footnotetext[2]{
Using the Bayesian approach described by the PDG \cite{pdg},
we obtain a 90\% C.L. upper limit of~7.9 for the number of 
$\TMG$ events. Inserting this value into eq.~(\ref{Beys0}) 
along with $\epsilon=12.0\%$, $\Omega=87.3\%$ gives 
${\mathcal{B}}(\TMG) < 4.8 \times 10^{-7}$. 
}

To take into account systematic uncertainties related to \~s$_{90}$, 
$(1+\Lambda)$ and $\kappa$ are varied by $\pm 1\sigma$ each.
This affects \~{s}$_{90}$ by $+0.06$/$-0.11$ events. 
The uncertainties in the functional forms of the background events 
are evaluated
by varying the most sensitive parameters by their evaluated errors.
The functional form is broadened or shortened by 1.2 or 0.8 times 
for $N^{\TT\GG}(\Minv,\DelE)$ and by 1.4 or 0.9 times for 
$N^{\mu\!\!\!/\mu \GG}(\Minv,\DelE)$, and 
their centers are shifted by $\pm$0.02~GeV for $N^{\TT\GG}(\Minv,\DelE)$ 
and by $\pm$0.015~GeV for $N^{\mu\!\!\!/\mu \GG}(\Minv,\DelE)$. 
These factors are about the largest that still give an 
acceptable fit to the background distributions.
The shift of the central value for the $N^{\TT\GG}(\Minv,\DelE)$ spectrum 
yields the largest effect of $\pm$0.2 events,
and the overall systematic uncertainty is evaluated as $\pm$0.3 events.
The stability of the result as the fit region is varied is checked by
extending the $\DelE$-$\Minv$ region to $\pm 8\sigma$; $s=0.0$ and $b=105.0$ 
are obtained, and \~{s}$_{90}$ varies by only 0.07 events. 

The systematic uncertainties on the detection sensitivity 
$2(\epsilon\Omega) N_{\TT}$ arise from the track reconstruction 
efficiency (2.0\%), photon reconstruction efficiency (2.8\%), 
selection criteria (2.2\%), luminosity (1.4\%), 
trigger efficiency (1.6\%), and MC statistics (0.8\%).
The total uncertainty is obtained by adding all of these components in 
quadrature; the result is~4.7\%.
This uncertainty is included in the upper limit on $\mathcal{B}(\TMG)$ 
following \cite{CLEO}\footnotemark[3].

\footnotetext[3]{
The quantity $1/(2(\epsilon\Omega) N_{\TT})$ is integrated assuming
a Gaussian distribution for $2(\epsilon\Omega) N_{\TT}$.
}

The angular distribution of the $\tau \to \mu \gamma$ decay
essentially depends on the LFV
interaction structure~\cite{okada},
and spin correlations between the $\tau$'s
at the signal and tagged sides must be considered.
To evaluate the maximum possible 
variation, $V-A$ and $V+A$ interactions are assumed;
no statistically significant difference in the $\DelE$-$\Minv$ 
distribution or in the efficiency is found compared to the case 
of the uniform distribution.

The incorporation of all systematic uncertainties increases the upper limit 
by 6.3\%, of which the effect of the $N_{BG}(\Minv,\DelE)$ uncertainty
dominates. 
As a result, the upper limit on the branching fraction is
\begin{equation}
\mathcal{B}(\TMG)<3.1 \times 10^{-7}~~~~ \rm{at~ 90\%~ C.L.}
\end{equation}
This result is lower than previous limits for this mode
and helps constrain physics beyond the Standard Model.

We wish to thank the KEKB accelerator group for their excellent
operation of the KEKB accelerator.
We are grateful to J.~Hisano, A.~Ilakovac and Y.~Okada 
for fruitful discussions.
We acknowledge support from the Ministry of Education,
Culture, Sports, Science, and Technology of Japan
and the Japan Society for the Promotion of Science;
the Australian Research Council
and the Australian Department of Education, Science and Training;
the National Science Foundation of China under contract No.~10175071;
the Department of Science and Technology of India;
the BK21 program of the Ministry of Education of Korea
and the CHEP SRC program of the Korea Science and Engineering Foundation;
the Polish State Committee for Scientific Research
under contract No.~2P03B 01324;
the Ministry of Science and Technology of the Russian Federation;
the Ministry of Education, Science and Sport of the Republic of Slovenia;
the National Science Council and the Ministry of Education of Taiwan;
and the U.S.\ Department of Energy.

\end{document}